\documentclass[12pt]{article}

\usepackage{epsfig}
\usepackage{amsmath}
\usepackage{amssymb}

\renewcommand{\Re}{\textrm{Re}\,}
\renewcommand{\Im}{\textrm{Im}\,}
\newcommand{\be}{\begin{equation}}
\newcommand{\ee}{\end{equation}}
\newcommand{\ba}{\begin{eqnarray}}
\newcommand{\ea}{\end{eqnarray}}
\newcommand{\nl}{\nonumber \\}

\newcommand{\OO}{{\cal O}}

\def\nB{n_\textrm{B}}

\def\lsim{\mbox{~{\raisebox{0.4ex}{$<$}}\hspace{-1.1em}
    {\raisebox{-0.6ex}{$\sim$}}~}}

\def\Nc{N_\textrm{c}}

\def\nr#1{(\ref{#1})}
\def\eqs#1{(\ref{#1})}
\def\la#1{\label{#1}}

\def\centerbox#1#2{\centerline{\epsfxsize=#1\textwidth \epsfbox{#2}}}

\def\g2{g^2_{d+1}}
\def\GR{G_\textrm{R}}
\def\GA{G_\textrm{A}}

\def\sg{\sqrt{{-}g}}
\def\sgamma{\sqrt{{-}\gamma}}

\begin{document}
\bigskip\bigskip
\bigskip\bigskip
\begin{center}
\textsc{\huge Hydrodynamic}
\end{center}

\begin{center}
\textsc{\huge Long-Time Tails}
\end{center}
\begin{center}
\textsc{\huge From }
\end{center}
\begin{center}
\textsc{\huge Anti de Sitter Space}
\end{center}
\bigskip
\bigskip
\bigskip
\centerline{\textsc {\large Simon Caron-Huot
\footnote{sicah@physics.mcgill.ca.
New address: Institute for Advanced Study, Princeton, NJ 08540 USA.}
\quad \large Omid
Saremi}\footnote{omid@physics.mcgill.ca}}
\bigskip
\bigskip
\centerline{\it{Ernest Rutherford Physic Building}}
\centerline{\it{McGill University }} \centerline{\it{Montreal,
QC}}\centerline{\it{Canada H3A 2T8}}

\begin{abstract}

For generic field theories at finite temperature,
a power-law falloff of correlation functions of conserved
currents at long times is a prediction of non-linear
hydrodynamics.
We demonstrate, through a one-loop computation in Einstein gravity in
Anti de Sitter space,
that this effect is reproduced by the dynamics of black hole horizons.
The result is in agreement with the gauge-gravity correspondence.
\end{abstract}
\vfill
\section{Introduction}
The AdS/CFT correspondence, or more broadly the notion of holography,
has provided us with a unique window into the dynamics of certain
strongly coupled field theories \cite{maldacena,GKP,Witten}. Viewed differently,
the boundary field theory can be seen as a realization of
a consistent quantum theory of gravity in an asymptotically
Anti de Sitter (AdS) space. The finite temperature is easily implemented
in AdS/CFT by introducing a black hole event horizon into the bulk.
A black hole horizon plays the role of an infrared
cutoff beyond which thermal fluctuations prohibit probing the vacuum.
A particularly simple limit of the AdS/CFT correspondence
is to focus on the low momentum/frequency regime where one
expects quantum field theories to exhibit universal behavior dictated
by hydrodynamics. More precisely,
field theories at finite temperature and in the low momentum and frequency
regime possess correlation functions
with a universal structure fixed by hydrodynamics. Reproducing
this structure from a weakly curved gravity background in asymptotically AdS space can
be regarded as a test for AdS/CFT. This is so because the field theory
side of the duality is strongly coupled even in the hydrodynamic
limit. Through the AdS/CFT dictionary, operators in the boundary
theory including fluctuations in the conserved quantities are mapped
into various bulk gravity modes. Exploring near-equilibrium phenomena
in the boundary field theory is holographically dual to studying
perturbation theory of certain black hole spacetimes.

In recent years, much effort has been devoted to
reproducing \emph{linearized hydrodynamics} and diffusion physics
in the context of AdS/CFT.
For instance, various transport coefficients
have been calculated in different boundary theories
and thermal ensembles  \cite{kovtunss}.
Some attention has also been paid to the hydrodynamic derivative expansion,
such as second order hydrodynamics, still linear in fluctuations \cite{stephanovetal}.

Non-linear hydrodynamics, on the other hand,
is host to numerous interesting phenomena associated with the non-linearities
in fluid dynamics, like turbulence.
In the context of holography,
it would be extremely interesting to see if there
exists a gravitational and/or stringy dual to, for instance,
a turbulent flow.

Reproducing non-linear terms in the hydrodynamics expansion
from gravity has been discussed in \cite{Shiraz}.
Having said that it is important to emphasize that in the context of AdS/CFT, \emph{phenomena} associated with non-linear hydrodynamics have
remained largely unexplored.
There are many interesting phenomena associated with
non-linearities in hydrodynamics, among them the so-called
\emph{long-time tails} which are the focus of this contribution.

Imagine $j^{\mu}=(\rho,j^a)$ to be a conserved current,
where $a,b,\ldots$ denote the spatial coordinates.
Consider the zero-momentum
self auto-correlator of $j^a$,
namely
 $\int d^{d-1}x \langle j^{a}(t,x)j^{b}(0,0)\rangle$,
where $d$ is the field theory spacetime dimension.
Recall the constitutive relations for $j^a$ in the hydrodynamic regime
\be
j^a \simeq -D\nabla^a \delta \rho + 
   \delta\rho~\delta u^a + \ldots ,
 \label{const1}
\ee
where $D$ is the diffusion constant.
The first term detects spatial inhomogeneities and
is diffusive, while the second term is convective, proportional
to the local flow velocity $u^a$. Omitted terms are suppressed
at long-wavelengths by additional derivatives or are cubic or higher in fluctuations.
At zero-momentum, only the second
non-linear term in \nr{const1} contributes.
It involves long-lived hydrodynamic fluctuations
(diffusive modes and sound waves), which
decay on diffusive time scales  $t\sim k^{-2}$ at wavenumber $k$.
This generates a power law
tail in the correlator
with an exponent which is a characteristic of diffusive physics
\begin{eqnarray}\label{long_time}
\int d^{d-1}x~\langle j^a(\tau,x)j^b(0,0)\rangle \propto \frac{\delta^{ab}}
  {\tau^{\frac{d-1}{2}}}.
\end{eqnarray}
The numerical prefactor in (\ref{long_time}) is
exactly calculable within the hydrodynamics framework
and the topic will be reviewed below \cite{ltt}.
Notice that this phenomenon originates from statistical
fluctuations around the equilibrium state so it
should be suppressed by the entropy density $s$.
For instance, in a deconfined gauge theory plasma
$s\sim N_c^2$, where $N_c$ is the number of colors,
 the long-time tail will be an $\OO(1/N_c^2)$ effect. According to the AdS/CFT dictionary,
this is mapped\footnote{Wherever a gravitational dual is
  available.} to a one-loop effect
in the bulk AdS quantum gravity.  This was noted sometime ago
by Kovtun and Yaffe \cite{KovtunYaffe}.
The tail \nr{long_time} cannot be seen in the classical gravity approximation.

Confirming the existence and reproducing this long-time tail
from a one-loop bulk gravity computation is the subject
of the present contribution. Interestingly, classical statistical fluctuations in the boundary theory
will be mapped to quantum gravity fluctuations in the bulk.


Quantum gravity is not a renormalizable theory so one might worry
about the prospects of performing a one-loop computation within its framework.
However, because we are dealing with an infrared
phenomenon, our final result will not depend on an ultraviolet completion.
The relevant one-loop integrals will be ultraviolet safe.
For precisely the same reason, stringy corrections will play no role; pure Einstein gravity and Yang-Mills theory in Anti de Sitter space will suffice.

It turns out that it is physically motivated to organize the one-loop computation
in a causal manner such that it does not rely on the physics beyond the black hole event horizon. We come
back to this point later in the text.

The organization of this paper is as follows.
We begin by reviewing the hydrodynamics predictions in section
\ref{sec:hydro}.
We take time to lay the formal grounds of our work throughout the
section \ref{sec:formal}.
At the same time, we clarify why on an intuitive level our
results are totally expected.
The reader familiar with the background material could jump directly to the
section \ref{sec:start}, perhaps after reading subsection \ref{sec:notation}.
Pre-requisites to compute the one-loop amplitude are discussed
in section \ref{sec:prereq}. It includes material ranging from computation
of the vertices to gauge-fixing.
Section \ref{finalsection} gathers these ingredients. Our paper closes with concluding remarks in section \ref{sec:conc}.
\section{Review of hydrodynamic predictions}
\label{sec:hydro}
We now sketch the calculation of the long-time tails in hydrodynamics.
For details, we refer the reader to \cite{ltt} which contains references to
experiments (see also a nice and more recent account of the topic by Kovtun and Yaffe,
in the context of relativistic fluids \cite{KovtunYaffe},
which we will be following closely here).

In a volume $V$, there are $sV$ independent
effective degrees of freedom where
$s$ is the entropy density.  By the central-limit theorem,
fluctuations of infrared modes are small and Gaussian-distributed. These fluctuations are fluctuations in conserved quantities.
In particular, their equal-time correlators are set by thermodynamics,
while their dynamics is governed by linearized hydrodynamics.
At arbitrary time separation the full two-point functions, for conserved charges and local fluid velocity, are given by \cite{ltt,KovtunYaffe}\footnote{
  In quantum field theory, the appropriate correlator is the symmetric
  correlator $V^{-1}\frac12 \langle \left\{\rho(t,k),
  \rho(0,-k)\right\}\rangle$, which is manifestly real \cite{KovtunYaffe}.
}
\ba\label{2ptfunc}
V^{-1}\langle \rho(\tau,k)\rho(0,-k) \rangle &=& \Xi e^{-Dk^2|\tau|},\\\nonumber
V^{-1}\langle u^i(\tau,k)u^j(0,-k) \rangle &=& \frac{T}{\epsilon{+}p}\left[
(\delta^{ij}{-}\frac{k^ik^j}{k^2}) e^{-\gamma_\eta k^2 |\tau|}
 + \frac{k^ik^j}{k^2} e^{-\frac12 \gamma_s k^2|\tau|}\cos(kc_{s}\tau)
\right]\!\!,
\ea
where $\Xi$ is the charge susceptibility, $c_s$ is the speed of sound,
$\gamma_{\eta}=\eta/(\epsilon+p)$ and
$\gamma_{s}=\zeta/(\epsilon+p)+2(d-2)\gamma_{\eta}/(d-1)$.

Hydrodynamics at finite temperature is to be viewed as an effective
statistical field theory.
In $d>(2{+}1)$ space-time dimensions,
hydrodynamics' canonical coupling constant has negative mass dimension.
Therefore, the theory becomes weakly-coupled in the infrared \cite{ltt,hydroIR}.
For this reason, a one-loop treatment becomes exact in the long-time limit.%
 \footnote{
 In $(2{+}1)$-dimensions, the canonical coupling becomes dimensionless and
 quantities run
 logarithmically with scale. For instance,
 the effective viscosity becomes large logarithmically
 in the infrared. A Kubo formula for the viscosity does not exist.
 Nevertheless, the $\beta$ function for the effective loop coupling
 turns out to be positive (at least for incompressible fluids), causing it to
 flow to a trivial fixed point in the infrared. This allows precise predictions to be made on long-time behavior
 \cite{hydroIR}.  }
Such a one-loop treatment amounts to approximating
\nr{const1} by its quadratic term
\be
\int d^{d-1}x~ \langle j^a(\tau,x) j^b(0) \rangle \simeq
 \int d^{d-1}x\, \langle \rho(\tau,x) u^{a}(t,x)\, \rho(0) u^{b}(0)\rangle.
\ee
This can be evaluated using the Gaussian two-point functions \nr{2ptfunc}
\be
 \langle \rho(\tau,x) u^{a}(\tau,x)\, \rho(0) u^{b}(0) \rangle
   = \langle \rho(\tau,x) \rho(0)\rangle
 \,\langle u^a(\tau,x) u^b(0)\rangle.
\ee
One finds \cite{ltt,KovtunYaffe}
\begin{subequations}
 \label{hydro}
\ba
\int d^{d-1}x ~\langle j^a(\tau,x) j^b(0) \rangle =
 \frac{T\Xi\delta^{ab}}{\epsilon+p}\frac{d-2}{d-1}
  \frac{1}{\left[4\pi(D{+}\gamma_{\eta}) |\tau|\right]^{\frac{d-1}{2}}},
 \label{hydro1}
\ea
which holds up to terms suppressed by higher inverse powers of $t$.
Here we have straightforwardly generalized
the results of \cite{KovtunYaffe}
to $d$ space-time dimensions.

Similar long-time tails are present in the stress tensor correlators.
Consider the correlator $\int d^{d{-}1}x \langle t^{xy}(t,x)t^{xy}(0)\rangle$.
There exists a second order non-linear term in the constitutive
relations for the stress tensor analogous to (\ref{const1})
with $t^{xy}\simeq (\epsilon{+}p) \delta u^x \delta u^y$.
One finds \cite{KovtunYaffe}
\be
 \int d^{d{-}1}x~\langle t^{xy}(\tau,x)t^{xy}(0)\rangle=
 \frac{T^2}{(d-1)(d+1)}\left[
  \frac{1}{(4\pi\gamma_s |\tau|)^{\frac{d-1}{2}}}
 +\frac{d^2-2d-1}{(8\pi\gamma_\eta |\tau|)^{\frac{d-1}{2}}}
 \right]. \label{hydro2}
\ee
\end{subequations}

\section{Long-Time tails from one-loop AdS gravity}\label{start}
\label{sec:formal}

Our goal will be to reproduce \nr{hydro1} and \nr{hydro2}
from a bulk one-loop gravity computation in AdS.  We find the
bulk diagrams which are dual to the hydrodynamic processes. We begin by
describing the setup.
\subsection{Setup and notations}  \label{sec:notation}
We use lower case Latin letters $i,j,\ldots=1\ldots d-1$ to denote the
spatial coordinates. The Greek letters are reserved for both temporal and spatial coordinates and run
over $0,1\ldots d-1$. Upper case Latin letters are used for the entire spacetime indices. The radial index will be referred to as ``$r$''. The superscript ``$(0)$'' will mean ``background value''. The bulk action we consider has the following form
\begin{eqnarray}
S_{\textrm{bulk}}&=&\frac{1}{16\pi G_{N}} \label{action}
 \int d^{d+1}x\sg(R-\frac{d(d-1)}{\ell^2})-
\frac{1}{4g^2_{d+1}}\int d^{d+1}x \sg F^2\\\nonumber
&&
+\frac{1}{8\pi G_{N}}\int d^{d}x \sqrt{{-}\gamma}K.
\end{eqnarray}
The background $AdS_{d+1}$-Schwarzschild black hole solves the equations of motion associated
to \nr{action}
\begin{equation}
 ds^2=g^{(0)}_{\mu\nu}dx^{\mu}dx^{\nu}=
 \left( \frac{4\pi T\ell}{d}\right)^2 dr^2
    +p(r)(-q(r)dt^2+\delta_{ij}dx^{i}dx^{j}), \label{metric}
\end{equation}
where
\begin{equation}
 p(r)= \cosh^{\frac{4}{d}}(2\pi T r),\quad
 q(r)=\tanh^2(2\pi T r).
\end{equation}
In these coordinates, the black hole horizon is located at $r=0$
and $T$ is its Hawking temperature.
For simplicity, we will work in units where $4\pi T\ell/d=1$.
The chemical potential is turned off so the gauge fields vanish in
the background.\footnote{Even pure gravity in Schwarzschild
 Anti de Sitter space has
 enough structure to exhibit the long-time tail phenomenon.}
In particular for $d=4$, this background is dual to strongly coupled
$\mathcal{N}=4$ super Yang-Mills theory at
temperature $T$ and vanishing R-charge
chemical potential \cite{Witten}.

For the bulk gravitational perturbations
$g_{\mu\nu}-g^{(0)}_{\mu\nu}=h_{\mu\nu}(r,x^{\mu})$,
we adopt the Gaussian normal gauge (spacelike axial gauge)
suitable in holographic settings
\begin{eqnarray}\label{gauge1}
h_{r\mu}=0, \quad  h_{rr}=0.
\end{eqnarray}
A suitable gauge choice for the gauge fields is
\begin{equation}\label{gauge2}
A_{r}=0.
\end{equation}
There will be still some residual gauge freedom rigid along the $r$ direction.%
\footnote{ This residual gauge freedom
  will eventually make a come back and haunt us,
however this will not be a major concern.}
\subsection{Schwinger-Keldysh Formalism}
\label{sec:keldysh}
Our aim is to discuss real-time physics in
the boundary CFT so we work directly in Minkowski signature.
For this we need the so called
Schwinger-Keldysh, or real-time formalism.
This allows for a description of expectation values with prescribed
initial states or density matrices which is appropriate for a thermal ensemble.
A comprehensive review of the method can be found
in \cite{Chou}.
In order to set our notations, here we include a lightning review
of the subject mostly based on \cite{herzog}.

\begin{figure}[t]
\centerbox{0.40}{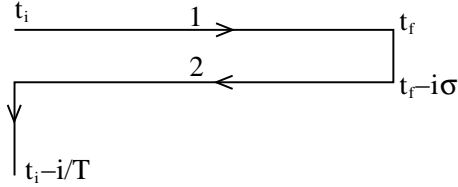}
\caption[Schwinger-Keldysh contour]
{
\label{fig:contour}
 Schwinger-Keldysh time contour for the finite temperature path integral.
}
\end{figure}
To accommodate operators with real time arguments,
the path integral time contour has to make an excursion to the real domain.
Since only the initial state is prescribed,
a doubled set of fields is required in the path integral:
one is to implement the forward time evolution up to the locations where operators are
inserted and the second one to implement backward time evolution back to the
time at which the initial state is prescribed. Averaging over a thermal ensemble of
initial states is done by adding a vertical Euclidean
part (with an extent $\beta=1/T$) to the time contour, with periodic
boundary conditions (antiperiodic for fermions).
This leads to a time contour depicted in Fig.~\ref{fig:contour}.
Denote the field values and sources
on the two horizontal parts
of the contour as
\be\label{doublet}
 \phi_{1}(t,\vec{x}) = \phi(t,\vec{x}),\quad \phi_{2}(t,\vec{x})=\phi(t-i\sigma),
 \quad J_{1}(t,\vec{x})=J(t,\vec{x}),\quad J_{2}(t,\vec{x})=J(t-i\sigma),
\ee
where the length of the vertical part of the contour is denoted by $\sigma$. Using this notation, one can write down a generating functional
for the real-time correlators as follows
\be
 Z[J_{1},J_{2}] = \int[\mathcal{D}\phi]
 \exp(iS + i\int_{t_i}^{t_f}dt\int d^dx J_{1}(x)\phi_{1}(x)
         - i\int_{t_i}^{t_f}dt\int d^dx J_{2}(x)\phi_{2}(x)),
\ee
where the Minkowski signature action $S=\int dtd^dx \mathcal{L}[\phi(x)]$
is evaluated along the contour.
The fields $J_{1}$ and $J_{2}$ are background sources for $\phi_{1}$ and its partner
$\phi_{2}$ respectively.
Due to the field doubling, the Schwinger-Keldysh
propagator is a $2\times2$ matrix given by second variation of the generating
functional
as usual
\be\label{sk}
 G_{ab}(x-y)=\frac{\delta ^2\ln Z}{\delta J_{a}(x)\delta J_{b}(y)}, \quad a,b=1,~2.
\ee
Note that $\sigma$ is arbitrary.
We will choose $\sigma{=}0$. At equilibrium, all these Green's functions are related to
the retarded and advanced Green's function through
Kubo-Martin-Schwinger (KMS) relations \cite{kapustagale} \footnote{Also known as
  fluctuation-disspation relations.}
\be \nonumber
 G_{11}= -i\Re\GR + (1+2\nB) \Im \GR,\quad
 G_{22}= i\Re\GR + (1+2\nB) \Im \GR,
\ee
\be
 \label{KMS}
 G_{12}= 2\nB\Im \GR,\quad
 G_{21}= 2(1+\nB)\Im \GR,
\ee
where $\nB$ is the Planck factor
\ba
 \nB(p)&=&\frac{1}{e^{\beta p^0}-1}, \nonumber
\ea
and
\ba
 \GR(x{-}y)\equiv i\langle [\phi(x),\phi(y)]\rangle \,\theta(x^0{-}y^0).
\ea
\subsection{ Schwinger-Keldysh Formalism in AdS/CFT }
\begin{figure}[t]
\centerbox{0.3}{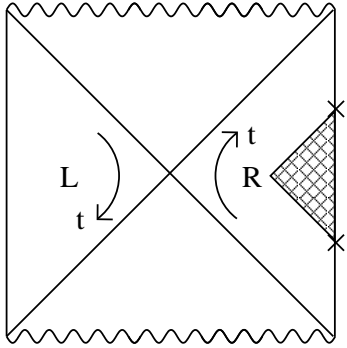}
\caption[Maximal Kruskal extension of AdS-Schwartzschild]
{
\label{fig:kruskal}
  The maximally extended geometry of an AdS-Schwartzshild black hole.
  The shaded region depicts the causal diamond associated to a causal boundary
  correlator. %
}
\end{figure}

\def\bdy{{\textrm{bdy}}}
\def\bulk{{\textrm{bulk}}}
Thermal QFT in the Schwinger-Keldysh
formalism and the Kruskal extension of a black hole
spacetime are connected \cite{IsraelThermo}. In a maximally extended eternal AdS black
hole spacetime as depicted in Fig.~\ref{fig:kruskal}, there are two causally disconnected
regions $R$ and $L$ with two AdS asymptotic regions. They provide
natural candidate regions for the doubled fields of the
Schwinger-Keldysh formalism to dwell on. Standard
AdS/CFT prescription \cite{Witten} for computing boundary
theory Euclidean amplitudes can be naturally extended to the case
where the bulk geometry is maximally extended. In this case the bulk partition function generates the Schwinger-Keldysh correlation
functions (\ref{sk}) at finite temperature $T$.\footnote{Provided the initial
  state for the Lorentzian black hole in the bulk is given by the
  Hartle-Hawking state \cite{maldacenaeternal}.}
This was accomplished by Herzog and Son \cite{herzog}.%
\footnote{A method for calculating
  retarded functions had previously been proposed by Son
  and Starinets \cite{sonstarinets} in planar coordinates.}

There is a useful formulation of this, which was proposed both by Skenderis and van Rees \cite{vanrees}
and by Son and Teaney \cite{sonteaney},
in which only the R-quadrant needs to be considered explicitly.
This is the formulation we shall use.
Let us first recall that the AdS/CFT correspondence equates the full
quantum gravity (string/M-theory) bulk path integral on AdS
with prescribed Dirichlet boundary conditions with
the generating functional of the boundary correlators
\begin{eqnarray}
 Z_\bdy[\phi_{\textrm{bdy}}]&=& \int[\mathcal{D}\phi]_
 {\phi|_\bdy{=}\phi_{\textrm{bdy}}} \exp(iS_{\bulk}[\phi]) \equiv Z_{\bulk}[\phi_\textrm{bdy}].
 \label{bdy}
\end{eqnarray}
Herzog and Son's approach identifies the path integral
on the Kruskal plane with boundary conditions on the fields and their
Keldysh partners in the quadrants $R$ and $L$ \cite{herzog}.
The boundary values of the bulk fields in quadrant $R$ and $L$,
$\phi^{bdy}_{1,2}$, are identified as sources in the Schwinger-Keldysh
formalism in the boundary theory.
What it is important here to note is general choices of $\sigma$ can be accomodated,
not only $\beta/2$ as was considered originally by Herzog and Son.
The geometry which we shall use is precisely Herzog and Son's geometry,
continued to $\sigma=0$.  This precisely brings quadrants L and R on top of each other.
This continuation was described precisely by Son and Teaney \cite{sonteaney}.

Skenderis and van Rees' obtain the same geometry by considering directly
the standard Schwinger-Keldysh contour depicted in
Fig.~\ref{fig:contour} (see Fig.~2 of \cite{vanrees} for a more precise
description), convoluted onto the black hole exterior region.
Independent of how one derives it,
this geometry concretely realizes the physical picture that
from the viewpoint of an exterior observer,
a black hole is described by a density matrix rather than an S-matrix.
This will allow us to study ordinary thermal field theory in the $R$-quadrant.

In the boundary theory, KMS conditions relate
all correlation functions to the retarded propagators.
The bulk theory in our limit is a ``field theory'' of interacting
gravitons and gauge particles propagating in the AdS black hole
spacetime.

The KMS relations also hold non-perturbatively in the bulk black hole spacetime,
for quantum fields in a fixed background.  This
has been confirmed long time ago by Gibbons and Perry
\cite{Gibbons}
\be
 G_{12}(r,r';p)= -i \nB(p) (\GR(r,r';p){-}G_{A}(r,r';p)),\quad\mbox{etc.},
 \label{KMSex}
\ee
where $G_{ab}(r,r';p)$ is the partial Fourier transform of the bulk
propagator function and $r,r'$ are two arbitrary radii.\footnote{
 Note that we use the invariant combination $-i(\GR{-}\GA)$
 in lieu of $2\Im \GR$ as in \nr{KMS}. The latter would be ambiguous
 in a mixed representation $(r,\omega)$, as it would depend on
 coordinate choice.
}
Relations \nr{KMSex} will play a key role.

Since here the metric itself fluctuates, we require an extension
of the Gibbons and Perry's result. We note that to
all orders, \footnote{In quantum field theory, the KMS relations hold
  non-perturbatively,
  however in quantum gravity it is not clear how to define ``bulk operators''
  non-perturbatively. Therefore, the meaning of \nr{KMSex}
  may be limited to a semiclassical expansion.}
the background AdS black hole geometry admits a time-like
Killing vector relative to which a
``temperature'' can be defined. The black hole Killing horizon will have constant surface gravity
everywhere on the horizon. Although, bulk loop effects can renormalize background
black hole's bare parameters (for instance, one expects $\delta
r_{H}/r_{H}\sim \mathcal{O}(1/N^2_{c})$),
they will not interfere with the KMS conditions.
This is so because we keep the (Euclidean) length of the time-like Killing vector at
the boundary fixed. This means that the computations are performed in the canonical
ensemble.\footnote{By equilibrium,
the periodicity of the Euclidean time circle is the same everywhere.}
Therefore, it is reasonable to expect the relation \nr{KMSex} is true to all loop orders in quantum gravity.

The retarded and advanced propagators are computable in the quadrant $R$.
In practice, the KMS relations remove any concern about the interior.
Put differently, to all orders,
any treatment involving the full Kruskal geometry
must agree with the predictions of an ordinary thermal field theory applied to the exterior
region, provided it is such that the equilibrium condition is respected.
\subsection{Dirichlet conditions for quantum fields in AdS}
In AdS/CFT we need to impose Dirichlet boundary condition \nr{bdy}
on the bulk path integral.
We follow Symanzik's treatment of quantum field theories on spaces with boundaries.
The boundary conditions are implemented by adding a boundary action
\cite{symanzik}. This method has been previously applied to certain
interesting problems.
The list includes studying Casimir effect \cite{symanzik} and critical
phenomena in the presence of boundaries \cite{DiDi}.
It will be manifest from this procedure
that the boundary correlators defined by
\nr{bdy} are in fact equivalent to the correlators of certain bulk
operators, i.e.,  the momentum conjugates of the
corresponding bulk modes.%
 \footnote{A similar observation was recently exploited by
  Iqbal and Liu \cite{IqbalLiu} for one-point functions in a background
  field, in a context closely related to ours. See also \cite{Papa}
 } 

We illustrate the idea for a bulk scalar field with the following action
\be
S_\textrm{bulk}=  -\frac{1}{2}\int_{r{<}r_c}  d^{d+1}x \sg
    (\nabla_\alpha\phi \nabla^\alpha \phi +2V(\phi)),
 \label{PIbulk}
\ee
where $r_{c}$ denotes the position of the regularized boundary.
Classically, the Dirichlet boundary condition
$\phi(r_c,x)=\phi_{\textrm{bdy}}(x)$
is obtained from varying the action if a suitable boundary term is added to the action
\be
S_\textrm{bdy} = \int_{r{=}r_c} d^dx \sgamma\,
\frac{\Pi_\phi(r_c,x)}{\sgamma} (\phi_{\textrm{bdy}}(x)-\phi(r_c,x)),
 \la{scalaraction}
\ee
where $\Pi_\phi=-\sg g^{rr}\partial_r\phi$ is the conjugate
momentum field for $\phi$, $\gamma_{\mu\nu}$ is the time-like induced metric
on the boundary and $\gamma$ is its determinant.
This suggests to use the path-integral
\be
 Z_\textrm{bulk}[\phi_{\textrm{bdy}}] = \int [\mathcal{D}\phi]
 e^{iS_\textrm{full}[\phi]}
 \label{pifull}
\ee
with $S_\textrm{full}=S_\textrm{bulk}+S_\textrm{bdy}$ as a replacement
for \nr{bdy}. Now $\phi(r_c,x)$ is treated as dynamical inside the path
integral.
To justify \nr{pifull} at the full quantum level,
one must further check, since the boundary condition
is obtained only as an equation of motion,
that $\phi(r_c,x)$ has no fluctuations around its mean
$\phi_{\textrm{bdy}}(x)$.
This can be readily verified at the free theory level --the combination
 $(\phi(r,x){-}\phi_\textrm{bdy}(x))$
converges to zero as a distribution when $r\to r_c$
in any two-point function\footnote{That is, when integrated
 against $x$-dependent test functions.} --More generally, this is proved to all orders in perturbation theory
around Eq.~(5.14c) of \cite{symanzik}.

Since the path integral
depends on $\phi_{\textrm{bdy}}$ only through $S_\textrm{bdy}$,
from \eqs{bdy}, (\ref{scalaraction}) and (\ref{pifull}),
we see that boundary theory correlators (obtained as variations
with respect to $\phi_\textrm{bdy}$) are equal to
the bulk theory correlators of $\Pi$'s
\be
 \langle \OO^{i_1}(x_1) \cdots \OO^{i_n}(x_n) \rangle_\textrm{bdy}
=
 \lim_{r_1,\ldots,r_n \to r_c} \langle \Pi^{i_1}(r_1,x_1)
   \cdots  \Pi^{i_n}(r_n,x_n)\rangle_\textrm{bulk}.
 \label{bdycorrel}
\ee
Now the Dirichlet boundary conditions are implemented automatically in the path integral. It should be clear that the
Schwinger-Keldysh indices $i_k$ are the same on both sides.
This identity will be verified in a concrete example
for two-point functions in section \nr{sec:bulkcalc}.
The bulk loop expansion of the right-hand side of \nr{bdycorrel}
gives rise to so called Witten diagrams \cite{Witten}.

In our set up, we will be dealing with bulk tensor
fields ($A_{M}$, $h_{MN}$) rather than scalars.
The construction here trivially generalizes to these cases. Since the
corresponding Yang-Mills and Einstein bulk
actions are also quadratic in derivatives.
In gravity, the conserved boundary currents and
stress tensor (densities) are
defined \cite{BrownYork,BalasubramanianKraus} as variations
of the on-shell action due to
variation of the boundary values of $g_{\mu\nu}$ and $A_\mu$
\begin{subequations} \label{currents}
 \begin{eqnarray}\label{stress}
 t^\mu_{~\nu} = 2\frac{\delta S_{\textrm{bulk}}}{\delta \gamma_{\mu\nu}}
  &=&\frac{\sgamma}{8\pi G_{N}}(\delta^\mu_{~\nu} K - K^\mu_{~\nu}),
 \\  \la{current}
  j^\mu =\frac{\delta S_\textrm{bulk}}{\delta A_{\mu}}&=&
  \frac{\sgamma}{\g2} F^{\mu r}, 
\ea
\end{subequations}
where a boundary limit $r\rightarrow r_{c}$ is understood. In our gauge
$K^\mu{}_\nu = \frac12 \gamma^{\mu\sigma}\partial_r g_{\sigma\nu}$
.
These operators, by definition, are
the (radial) momentum conjugate to the metric and gauge fields.%
  \footnote{Working with the tensor densities turns out to be more convenient
  than working with the tensors.}
They play the role of $\Pi_\phi$ in \nr{bdycorrel}.
Note that the bulk equations of motion imply their conservation.


The momentum conjugates $\Pi_\phi$ have $\delta$-function
contact singularities proportional to
$\delta(r_i{-}r_j)\delta^d(x_i{-}x_j)$ in their correlators. This forbids the $r\to
r_c$ limit to be taken directly inside \nr{bdycorrel}.
In general, additional contact terms proportional
to $\delta^d(x_i{-}x_j)$ and its derivatives
may be necessary to be added to the right-hand side of \nr{bdycorrel} for the
equality to hold.
Within bulk perturbation theory, these contact term ambiguities
as well as the \emph{bulk renormalization} of
\nr{bdycorrel} are discussed by Symanzik \cite{symanzik}.
These are distinct from the so called \emph{holographic renormalization}
issues, which arise when the limit $r_c\to\infty$ is taken
in keeping with the AdS/CFT prescription.
Additional counter-terms, intrinsic to the boundary metric and matter, may be necessary when the regulator $r_{c}$ is taken to infinity \cite{skenderisrenorm}.
This amounts to subtractions in (\ref{stress}) and (\ref{current}).
It is worth emphasizing that these counter-terms will
not affect our long-time tail discussion.
%
\subsection{Flow equations}
It will be useful to think of the operators in \nr{currents}
as extending into the bulk and not just as boundary operators.
That is, we consider more generally
\begin{eqnarray}\label{r-dep}
t^{\mu}_{\nu}=t^{\mu}_{\nu}(r), \quad j^{\mu}=j^{\mu}(r).
\end{eqnarray}
Note that at finite $r$, they lose their meaning as local current densities.\footnote{There is no ``local'' stress tensor associated to the gravitational field.}
We derive simple radial flow equations
for these currents in a very similar way to \cite{IqbalLiu}.

We only quote the flow equations we need.
As expected, it turns out that the operators (\ref{stress})
and (\ref{current}) have simple radial profiles in the
hydrodynamic limit. The following equation is easy to
prove from Einstein's equations
\begin{equation}\label{useful}
 \frac{1}{\sqrt{g}}\partial_{r}(\sqrt{g}K^{\mu}_{~\nu})={}^{(d)}R^{\mu}_{~\nu}
+\frac{d}{\ell^2}\delta^{\mu}_{~\nu}.
\end{equation}
Here ${}^{(d)}R^{\mu}_{~\nu}$ is the Ricci tensor of the induced
metric on the constant-$r$ slices.
Using Yang-Mills' equations of motion and
(\ref{useful}),
one obtains the following radial flow equations for the traceless part
of the stress tensor and the current fluctuations
\begin{subequations}
  \label{radialflow}
 \begin{eqnarray}
  \partial_r (\delta t^{\mu}_{~\nu})=
 \frac{1}{8\pi G} \delta(\sg {}^{(d)}R^{\mu}_{~\nu}) =
 0+\OO(\omega^2,q^2) \quad \mu\neq \nu,
 \\
  \partial_r (\delta j^\mu) =\frac{1}{\g2} \delta(\partial_\nu \sg
 F^{\nu \mu}) = 0+\OO(\omega^2,q^2),
\end{eqnarray}
\end{subequations}
where we have used the fact that ${}^{(d)}R^{\mu}_{~\nu}\propto \delta^\mu{}_\nu$
evaluated on the background. Also in momentum space,
$\delta [{}^{(d)}R^{\mu}_{~\nu}]=\mathcal{O}{(\omega^2,q^2)}$,
since $\mu$ and $\nu$ are tangential to the
constant-$r$ hypersurface.
It is worth noting that errors on the right-hand sides are not uniform in $r$.
As we will see later, due to the near horizon oscillations
in the field profiles $\propto r^{\pm i\omega/2\pi T}$, the right-hand sides in (\ref{radialflow}) are small only up to exponentially small radius
$r\sim e^{-2\pi T/\omega}$.

In sum, in the hydrodynamic regime the radial profiles of the
currents are simply constant over the bulk geometry expect for a an exponentially small region near the horizon.
Clearly, it makes sense to organize the bulk computation in terms of these variables.
\subsection{What to compute}
Computing a one-loop \emph{fluctuation} function
in a black hole background directly would be very difficult:
in following the past history of a fluctuation within perturbation theory,
we would have to inevitably cross
the black hole horizon in a finite proper time, rendering such a calculation
daunting. However, we have argued above that the details of the interior region
must be irrelevant at the end, as long as this region is treated in such a way
that the KMS relations are respected. What we will compute is the zero-momentum retarded two-point functions of the bulk operators
$j^x$ and $t^x{}_y$, at one-loop
\ba
 {}^{(1)}\GR^{xx}(\tau) &\equiv \int d^{d-1}x~{}^{(1)}
    \langle j^x (\tau,x,r\to\infty) ~j^x (0,r\to\infty) \rangle_\textrm{R}\,,
\nl
 {}^{(1)}\GR^{xy,xy}(\tau) &\equiv \int d^{d-1}x~{}^{(1)}
    \langle t^x{}_y (\tau,x,r\to\infty) ~t^x{}_y (0,r\to\infty) \rangle_\textrm{R}\,.
\ea

It should be completely obvious from \nr{bdycorrel} that the boundary thermal field theory inherits
the KMS relations from the bulk.
The long-time limit of the fluctuation function then follows from that
of the retarded function using
\be
\frac12 \int d^{d-1}x ~{}^{(1)}\!\langle \{ j^x(\omega,x),j^x(0)\} \rangle =
 (1 + 2\nB(\omega))~ \Im {}^{(1)}\GR^{xx}(\omega)
\ee
in frequency space.  There is a similar relation for $t^{x}{}_y$ correlators.

\begin{figure}[t]
\centerbox{0.50}{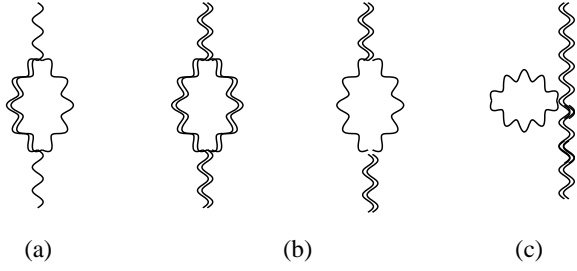}
\caption[Bulk loop diagrams]
{
 \label{fig:diagsg}
 Feynman diagrams contributing to (a) current correlator (b) stress
 tensor correlator. Four-point vertices (c) will be found to have negligible
 effects. Wavy lines are bulk Yang-Mills fields and double lines are
 gravitons.
}
\end{figure}
\subsection{Emergence of hydrodynamics}  \label{sec:start}
\begin{figure}[t]
\centerbox{0.45}{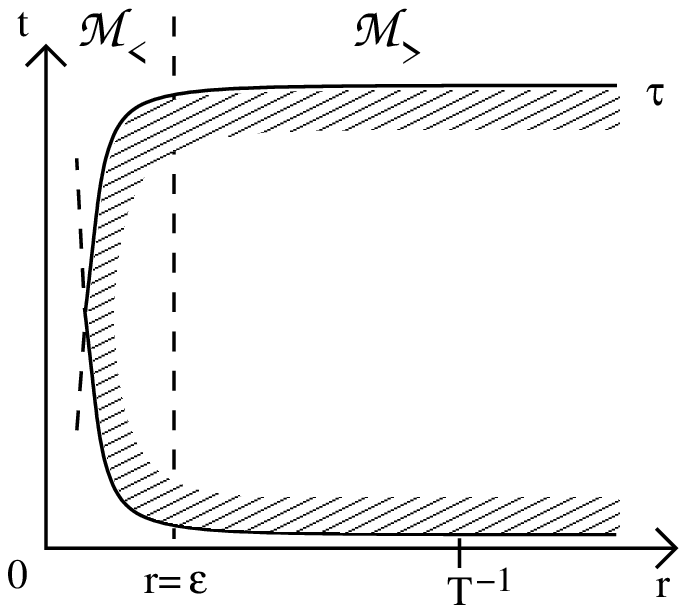}
\caption[Space-time structure of the diagram]
{
\label{fig:spacetime}
 Space-time structure of the integration region for the loop vertices (``half-almond'').
 The external wavefunctions force the vertices to localize in the shaded region.
 Hydrodynamics interactions are generated in
 region $\mathcal{M}_>$ when both vertices are at radii of order $T^{-1}$.
}
\end{figure}

The relevant Feynman diagrams are depicted in Fig.~\ref{fig:diagsg}.
Due to causality, the vertices will be restricted to lie inside the causal
diamond emanating from the insertion points at the boundary,
as depicted in Figs.~\ref{fig:kruskal} and \ref{fig:spacetime}.

In the $r-t$ plane, the integration region $\mathcal{M}$ for the two vertices
is the half-almond depicted in
Fig.~\ref{fig:spacetime}.
For the metric \nr{metric} its boundary can be described
precisely,
\begin{equation}
 t = \frac \tau2 \pm (r_{*}(r)-\frac \tau2), \quad
 r_{*}(r)= \frac{1}{2\pi T}\left[\tan^{-1}(y)-\tanh^{-1}(y)\right],
 \quad y= [\cosh(4\pi Tr)]^{-\frac12}. 
\end{equation}
where $\tau$ is the (long) time separation between the two boundary
insertions.
For $r\gg 1/(2\pi T)$,
the null geodesics are approximately horizontal lines, $t\approx 0$,
$t\approx \tau$.
In the $\tau\rightarrow \infty$ limit, in which we are interested in,
the tip of the half almond is approximated by
\begin{eqnarray}\label{tip}
 r_{0}(\tau)\approx\frac{\sqrt{2}}{2\pi T}e^{-\pi T \tau}.
\end{eqnarray}
Therefore, in this limit, the tip is exponentially close to the horizon.

The integration will not be dominated by the whole half-almond $\mathcal{M}$.
As will be seen in section \nr{external},
the external wavefunctions decay exponentially on a microscopic time
scale $T^{-1}\ll t$ as one moves vertically in the diagram.
This will localize the integration regions to the shaded area in
Fig.~\ref{fig:spacetime}, which has a thickness of order $T^{-1}$.
We will see this in more detail shortly.

In addition, it will turn out that not all radii equally contribute.
Heuristically, according to the scale-radius entry of the AdS/CFT dictionary,
one can deduce that the radius at which a bulk vertex is evaluated corresponds to its non-locality scale
in the boundary theory.
In particular, vertices near the tip of the almond would correspond
to interactions spread over a time interval $\sim \tau$, which may seem unexpected
since vertices in the hydrodynamics are local in time and instantaneous. They correspond to a spread over a microscopic time interval $\sim
T^{-1}$ (see Fig.~\ref{fig:magnifying} below).
With this in mind, we separate the bulk into two regions,
$\mathcal{M}_<$ and $\mathcal{M}_>$ at $r=\epsilon$,
with $\epsilon$ small.
Basically $\epsilon$ is chosen to separate the ``tip'' region of the half-almond
from the rest.  An adequate choice would be
\be
 \epsilon(\tau) \sim \frac{1}{T} e^{-\pi T\tau/2} \label{epsilonchoice}.
\ee
With this choice, the time separation between the vertices is large  and order of $\tau$, everywhere in $\mathcal{M}_>$.
An important finding will be that the contribution from $\mathcal{M}_<$
is exponentially suppressed in $\tau$.
The vertex integrations in region $\mathcal{M}_>$ will give
rise to the effective hydrodynamics interactions discussed in section
\ref{sec:hydro}.
This will be shown in sections \nr{sec:vertex} and \nr{sec:proof}.
This is how the one-loop boundary hydrodynamics emerges from the bulk gravity.

%
\section{Pre-requisites for computing the diagrams} \label{sec:prereq}
\subsection{External Wavefunctions}
\label{external}
Let us define the zero-momentum gauge field wavefunction $A_{x}=\psi_{\omega}(r)e^{-i\omega t}$.
The boundary to bulk propagator for $A_{x}$ and $h^{y}{}_{x}$ are just the
wavefunction for
these bulk fluctuations
satisfying  $\psi_{\omega}(r)=1$ when $r\rightarrow \infty$. The appropriate boundary conditions are retarded for the field sourced by the
field insertion at $\tau=0$ and advanced for the field sourced by the other one at time $t=\tau$.\footnote{This can be derived within the
  Schwinger-Keldysh formalism. See subsection \ref{sec:pert}.}
The retarded boundary condition implies
$\psi_{\omega}(r) \propto  r^{-i\frac{\omega}{2\pi T}}$
as $r\rightarrow 0$.
\def\FF{\mathcal{G}}

The linearized equations of motion for $A_x$ give
\begin{eqnarray}
 \frac{1}{\FF}\partial_{r}[\FF\partial_{r}\psi_{\omega}(r)]
 + |g^{tt}|\omega^2 \psi_{\omega}(r)=0,
 \label{scalarlikeeq}
\end{eqnarray}
where $\FF=\sqrt{-g}g^{xx}$.
The graviton field $h^y{}_x$ can be discussed similarly instead $\FF=\sqrt{-g}$. The wavefunction $\psi(r)$ with these boundary conditions is
\begin{equation}\label{wavefunction}
 \psi_{\omega}^\textrm{ret}(r)=\frac{{}_2F_{1}(-\frac{(1+i)}{2}\mathfrak{w},-\frac{(-1+i)}{2}\mathfrak{w};1-i\mathfrak{w};\tanh^2{(\pi
 T
 r)})}{{}_2F_{1}(-\frac{(1+i)}{2}\mathfrak{w},-\frac{(-1+i)}{2}\mathfrak{w},1-i\mathfrak{w},1)}\tanh^{-i\mathfrak{w}}{(\pi
 T r)},
\end{equation}
where $\mathfrak{w}=\frac{\omega}{2\pi T}$ and
\begin{eqnarray}
{}_2F_{1}(-\frac{(1+i)}{2}\mathfrak{w},-\frac{(-1+i)}{2}\mathfrak{w};1-i\mathfrak{w};1)=\frac{\Gamma(1-i\mathfrak{w})}{\Gamma(1+\frac{\mathfrak{w}}{2}-i\frac{\mathfrak{w}}{2})\Gamma(1-\frac{\mathfrak{w}}{2}-i\frac{\mathfrak{w}}{2})}.q
\end{eqnarray}

The general features of the time domain wavefunction (\ref{wavefunction}) are easy to obtain.
Note that the wavefunction (\ref{wavefunction}) has quasinormal poles
at \footnote{Similar computations have been performed in numerous previous works. See for example \cite{Berti}.}
\begin{eqnarray}
 \mathfrak{w}_{n}=n(\pm1 -i), \quad n=1,2, \ldots .
\end{eqnarray}
This indicates that $A_x$ fluctuations relax exponentially on a microscopic time scales as
the quasinormal modes are gapped and for all $n$ $\Im \mathfrak{w}_n\neq0$.
By construction, near the boundary the wavefunction is approximated by
$A_{x}(t)\approx\delta(t)$.
Near the horizon the dominant $r$-dependence
is due to the $r^{-i\mathfrak{w}}$ factor coming from expanding $\tanh$. From this we conclude that
\be
 A_x^\textrm{ret}(r,t) \to A_x^\textrm{ret}(t-r_{*}(r)),
\ee
which indicates that a given field profile falls into the black hole following
null geodesics.


An important property of $\psi(r,t)$
is that its time integral $\int dt\,\psi(r,t)=1$ is independent of $r$.
This follows from the $\omega\to0$ limit of \nr{wavefunction};
The only solution to \nr{scalarlikeeq} with $\omega=0$ which is
nonsingular at the horizon, is a constant.
The same is true for $h^y{}_x$.
These general properties are expected to hold for any AdS black hole.
%
\subsection{Radial and time integrations for vertices in $\mathcal{M}_>$}
\label{sec:vertex}
In this subsection, we work out the bulk gravity interaction vertices corresponding to
the non-linear hydrodynamics. We will show that the dominant vertices in the hydrodynamic limit can be written
in such way that they only involve the bulk operators
$t^{\mu}_{~\nu}(r)$ and $j^{\mu}(r)$ defined in the
previous section. This makes the holographic map between
the hydrodynamic diagrams contributing to the long-time tail and the
bulk quantum gravity amplitudes most transparent.
For the moment we only focus on the region $\mathcal{M}_>$, where
several approximations are applicable.
\subsubsection{Dominant fields in hydrodynamic regime}
First we must identify the bulk fluctuations which are long-lived.
We consider fields in Fourier space with four-momentum
components $(\omega,0,0,k)$ which are taken to be all small, compared to $T$. We
look for narrow peaks in correlation functions.
The radial flow equation \nr{radialflow} can be trusted in region $\mathcal{M}_>$ in this kinematic regime.

For the gauge fields, the transverse components do not exhibit
diffusive poles; they relax on microscopic scales so they can be ignored
in the long-time analysis.
The $A_t$ and $A_z$ components mix but because of Gauss' law,
\be\label{Frt}
 \omega F^{rt} = k F^{rz},
\ee
we have that $A_t/A_z\sim k/\omega\gg 1$ since $\omega\sim
k^2/T$ for a diffusive mode. Thus, $A_t$ dominates long-time correlations.\footnote{Note that this in only true in $\mathcal{M}_>$ away from the horizon. See subsection \ref{sec:proof}.}
Stress tensor correlators in the shear channel behave similarly,
with $h_{tx}$ dominating over $h_{zx}$.

Sound excitations involve the fluctuations $h_{tt}$, $h_{tz}$, $h_{zz}$ and
$h_{xx}+{\ldots}+h_{zz}$ (spatial trace) \cite{sonstarinetssound}. They are all of the same
order, since $\omega/k\approx c_s=\OO(1)$ for sound waves. Note that all diagonal components
of the spatial metric $h_{ij}$ are involved.
In fact, the on-shell fluctuations of $h_{zz}$ are equal to those
of $h_{xx}=h_{yy}=\ldots$ up to subleading viscous corrections (which are further suppressed by more powers of momentum). The spin-2 fluctuations can be neglected. This will be important later on. It can be shown as follows.
The radial flow equation \nr{radialflow} for the spin-2 part of $t^{ij}$ is initialized by the boundary
hydrodynamic data. The initial data is provided by the linearized constitutive relation for the stress tensor
$\delta t_\textrm{bdy}^{ij}\simeq \delta p\delta^{ij} + \eta \nabla^{(i} u^{j)}+\OO(\nabla^2)$.
The observation is that the spin-2 part is a viscous effect and therefore, suppressed by the momentum.
Linearizing the definition \nr{currents}, this demonstrates that the spin-2 component of $\partial_r h^i{}_j\propto \delta t_\textrm{bulk}{}^{i}{}_j$
is also small in the bulk. One can evolve this data into the bulk using (\ref{radialflow}).
Since the boundary value of $h^{i}{}_{j}$ is itself zero, its spin-2 component will remain subdominant in the bulk.
These estimates are outlined in Table~\ref{tab:modes}.
\begin{table}
\begin{tabular}{l||l||l}
Mode & Frequency scale & Fields turned on \\ \hline \hline
Charge diffusion poles & $\omega\propto k^2/T$ & $A_t\gg A_z\sim (q/T) A_t$
\\
Shear poles & $\omega\propto k^2/T$ & $h_{tx} \gg h_{zx}\sim (q/T) h_{tx}$
\\
Sound poles & $\omega\propto k$     &
 $h_{tt}\sim h_{tz} \sim h_{zz}\approx h_{xx}=h_{yy}$ (no spin-$2$)\\
\end{tabular}
\caption[Hydrodynamic modes]
{
 \label{tab:modes}
 Dominant bulk fluctuations in the long-term limit, with $z$ being
 parallel to the spatial  momentum.
}
\end{table}
\subsubsection{Bulk interaction vertices: current $j^x$}
Our first step is to couple the
boundary operator $j^x$, or equivalently the bulk perturbation
$\delta A_x$ sourced by it,
to bulk three-point vertices.
We concentrate on the ``future'' vertex near $t=\tau$.
We begin with the gauge-gauge-graviton vertex. This vertex is a third variation of Yang-Mills' (or Maxwell's) action
\ba
&& V_{A-A-g}= \delta_g \delta_{A}\delta_{A} S_{YM}
=-\int d^{d+1}x
\,\delta A_x
\,\partial_M [\delta_{g}\delta_{A}
(\frac{\sg F^{xM}}{\g2})]
\nl
&=& \int d^{d+1}x \,\frac{\delta A_x}{\g2}
 \partial_M [ \sg(
    \delta F^{\mu M}h^x{}_\mu
   +\delta F^{x\mu} h^M{}_\mu
   -\frac12 \delta F^{xM} h^\sigma{}_\sigma)].
 \label{vertexg1}
\ea
Notice the particular placement of the partial derivatives; this is a natural choice
as it exhibits the boundary current as an integral over a ``bulk current''. When we estimate region $\mathcal{M}_<$ below, it is imperative to
consistently keep up with the same convention. \footnote{Any other choice would be related to this one by a surface term. Notice that, by virtue of causality and smoothness of the external
wavefunctions, and the boundary conditions at infinity, such
surface terms associated to the integration-by-parts vanish.}

In the region $\mathcal{M}_>$, the individual terms appearing in the sum over indices
in the second term are either $k$-suppressed compared to other terms or vanish by our choice of gauge, so we safely ignore them.
The loop fields obey Maxwell's equations, up to contact terms.
Such terms would have the effect of pinching two vertices together,
which is exponentially suppressed in $\mathcal{M}_>$ due to the external wavefunctions.
It will not affect the long-time tail physics. In essence, a contact terms would turn two cubic vertices into
a four-point vertex, which is negligible in $\mathcal{M}_>$ for the same reason.
Thus the internal lines will be taken to be on-shell.
Keeping only the contribution from $M=r$ (the other indices being $k$-suppressed),
the remaining two terms reduce to
\be
V_{A-A-g}=\int d^{d+1}x \,\delta A_x \frac{\sg}{\g2}
    \left[\delta F^{\mu r} \partial_r h^x{}_\mu
   -\frac12 \delta F^{xr}\partial_r h^\sigma{}_\sigma \right]. \label{dummyggA}
\ee
According to Table \ref{tab:modes}, fluctuations in $F_{tr}$ dominate
over those of $F_{ir}\sim (k/T)F_{tr}$, which makes the second term
in the bracket sub-leading.
In any case, this term could not produce any long-time tail
since it mixes a sound mode with a diffusion mode, which do not overlap
in spacetime.
Thus we keep only the first term with $\mu=t$
\be
 V_{A-A-g}=\int d^{d+1}x \,\delta A_x \frac{\sg \delta F^{tr}}{\g2} \partial_r h^x{}_t.
\nonumber
\ee
Linearizing the definitions \nr{current} and \nr{stress} this may be written
as
\be
V_{A-A-g}=16\pi G \int d^{d+1}x \,\delta A_x  \frac{g^{xx}}{\sg |g^{tt}|}
\delta j^t(r)  \delta t^t{}_x(r).
\nonumber
\ee
The profile for $\delta A_x$ has a time extent of order $T^{-1}$ which
is much smaller than the long time $\tau$, which separates the vertices.
Since its time integral is 1, it can be approximated as a $\delta$-function
at the cost of $\OO(1/\tau)$ corrections
\ba
V_{A-A-g}&=& \int d^{d-1}x \,\delta j^t(\tau) \,\delta t^t{}_x(\tau)
  \int \frac{dr 16 \pi G g^{xx}}{\sg |g^{tt}|}  + \OO(1/\tau)
\nl
&=& \int d^{d-1}x \frac{\delta j^t(\tau) \,\delta t^t{}_x(\tau)}{\epsilon{+}p},
\label{vertexg2}
\ea
the error being a relative error.
The last line follows from the integral representation of
the inverse momentum susceptibility $1/(\epsilon{+}p)$ \cite{IqbalLiu}.

For theories with non-Abelian global symmetries the associated bulk gauge fields
are non-Abelian, and we need to consider the associated gauge-gauge-gauge
vertices.  Keeping only terms
with one radial derivative, using rotational invariance, one finds only
products $\delta A^x\partial_r \delta A^t$ or $\delta A^t\partial_r \delta A^x$,
which are sub-leading compared to \nr{vertexg2} in the hydrodynamic
expansion. Therefore, these interactions are sub-leading (at zero chemical potential).

In sum, in the region $\mathcal{M}_>$ we find that at zero spatial momentum
\be
 j^x(\tau)\simeq \frac{\delta j^t(\tau) \,\delta t^{tx}(\tau)}{\epsilon{+}p} +\OO(1/\tau).
 \label{vertexj}
\ee
A similar result holds for the ``past'' vertex near $\tau=0$.
\subsubsection{Bulk interaction vertices: stress tensor $t^{xy}$}
We repeat the same procedure for the bulk field $h{}^y_{~x}$ dual to the boundary $t^{xy}$ fluctuations.
The corresponding bulk interaction is a graviton-graviton-graviton vertex. This vertex is a third variation of the Einstein-Hilbert action
\ba
 && V_{g-g-g}=\delta_{g}\delta_{g}\delta_{g}S_{EH}= \frac{1}{16\pi G}\delta^3\int d^{d+1}x \sg(R+d(d-1)/\ell^2) \nl
 &=&{-}\frac{1}{8\pi G} \int d^{d+1}x h_{xy} \delta^2 \sg g^{x\mu} g^{y\nu}
 \left[ R_{\mu\nu} -\frac12 g_{\mu\nu}R -\frac{d(d{-}1)}{2\ell^2} g_{\mu\nu}\right].
 \la{dummygrav}
\ea
Similar to the $j^{x}$ case, the position of the derivatives in
\nr{dummygrav} is important. The double variation represents the insertion of the two gravitons running in the loop.
Notice there is a factor of two
coming from the fact that both $h_{xy}$ and $h_{yx}$
are sourced by the same profile.

We evaluate the double variation in \nr{dummygrav} using Leibniz's rule. The quantity in the bracket is the equation of motion of the background
and it vanishes when no variation acts upon it. When only one variation acts upon it this yields the equation of motion
of the linearized perturbation, which vanishes up to contact terms. These are negligible in $\mathcal{M}_>$ as explained in the previous subsection.
The equation \nr{dummygrav} reduces to
\be
 V_{g-g-g}={-}\frac{1}{8\pi G}
\int d^{d+1}x \sg h^{xy} \delta^2 \left[ R_{xy}-\frac12 g_{xy}R -
 \frac{d(d{-}1)}{2\ell^2}g_{xy}\right]. \nonumber
\ee
The last term manifestly has no second variation. Using equations of motion of the background plus those of the linearized
perturbations, one can show that $\delta^2
(g_{\mu\nu}R)=g_{\mu\nu}\delta^2 R$.
Therefore, the second variation of the second term vanishes since
there are no background off-diagonal metric components. We will have
\be
 V_{g-g-g}={-}\frac{1}{8\pi G}\int d^{d+1}x \sg h^{xy} \delta^2 R_{xy}.
\nonumber
\ee
At this stage we can begin to make hydrodynamic
approximations.  This means we only keep terms
in $R_{xy}$ which contain two radial derivatives
\be
 V_{g-g-g}={-}\frac{1}{8\pi G}\int d^{d+1}x \sg h^{xy} \delta^2 \left[
  2K_{\sigma x} K^\sigma{}_{y} - K^\sigma_\sigma K_{xy} -\frac12 \partial_r^2 g_{xy}\right].
\ee
The last term manifestly has no second variation.
Following Table \ref{tab:modes}, we keep only perturbations
of diagonal components of the metric and of $h^{ti}$.
At this point the fact that long-lived spin-2 excitations are absent proves important;
by rotational invariance, it is impossible to couple
$h_{xy}$ to two diagonal modes. It is also impossible to
have only one $h^{ti}$ insertion because of parity in the index $t$.
Therefore, only the terms whose index structure allow for two $h^{ti}$
insertions are kept
\be
 V_{g-g-g}=\frac{1}{4\pi G}\int d^{d+1}x \sg h^{xy} |g_{tt}|
\delta K^t{}_x \delta K^t{}_y.
\nonumber
\ee
Finally, utilizing $\int dt~h^y{}_x\approx 1$ plus the radial flow equations
(\ref{radialflow}) as above,
we obtain
\ba
V_{g-g-g}&=& \int d^{d-1}x \,\delta t^t{}_x \,\delta t^t{}_y \int dr \frac{16\pi G g^{xx}}
{\sg |g^{tt}|}  + \OO(1/\tau)
\nl
&=& \frac{1}{\epsilon{+}p} \int d^{d-1}x \,\delta t^t{}_x \,\delta t^t{}_y,
 \label{vertexh2}
\ea
the radial integral being as above.

The graviton-gauge-gauge vertex is sub-leading. The reason is, by rotational invariance (and keeping
only terms with two radial derivatives),
this can only couple to $F^{xr}F^{yr}\sim j^x j^y$ which according to
Table~\ref{tab:modes} is subdominant compared to \nr{vertexh2}. This implies that the graviton-gauge-gauge vertices do not contribute to long-time
tails at zero chemical potential.
From this, at zero spatial momentum
\be
 t^{xy} \simeq \frac{\delta t^{tx} \,\delta t^{ty}}{\epsilon{+}p} + \OO(1/\tau).
 \label{vertext}
\ee
\subsection{Bulk-Bulk propagators}
\label{sec:bulkcalc}
In the preceding subsection, we have ignored $\omega$ and $k$ dependent corrections in \nr{radialflow}.
This amounts to equating the bulk-bulk propagators of the currents in the loop to their boundary limits.
These boundary correlators can be found in \cite{sonstarinets,sonstarinetssound}
\ba
 \GR^{tt}(r,r';\omega,k) &\approx&
   i\Xi \frac{Dk^2}{\omega + iD k^2},\nl
 \GR^{ti,tj}(r,r';\omega,k) &\approx&
  i\gamma_\eta w \frac{(k^2\delta^{ij}-k^ik^j)}{\omega + i\gamma_\eta k^2}
\nl &&
-\frac{k^ik^j}{k^2} \frac{ \epsilon \omega^2 + p c_s^2 k^2}
   {(\omega+i\frac12 \gamma_s k^2)^2-c_s^2k^2},
 \label{bulk2pt}
\ea
where for $\mathcal{N}=4$ SYM theory, $D=\frac{1}{2\pi T}$,
$\Xi= \frac{\Nc^2T^2}{8}$, $p=\frac13\epsilon=\pi^2N_c^2T^4/8$ and
$r,r'$ are two arbitrary radii in $\mathcal{M}_>$.
Note that the imaginary part of these retarded functions times $2T/\omega$ reproduces the Fourier transform of the hydrodynamic expressions for the fluctuation functions \nr{2ptfunc}.

To consolidate this intuition and validity of the approximation \nr{bulk2pt}, here
we compute the bulk-bulk propagators in the hydrodynamic limit.
As we will see in this section, even at small $\omega$ and $k$ expressions \nr{bulk2pt} will break down at exponentially small $r, r'$ in
$\mathcal{M}_{<}$ due to oscillations.
Furthermore, if one applies the radial flow equations to a two-point function,
the right-hand side of \nr{radialflow} will generically contain a
contact term. This leads to the appearance of contact terms $\propto \delta(r-r')$ in \nr{bulk2pt}. This is generically
expected to be the case for correlator with ``momentum'' variables.
In sum, equation \nr{bulk2pt} holds at small $\omega,k$ (or long times),
where $r$ and $r'$ are not-exponentially-small. Also contact terms are expected in the expressions
\nr{bulk2pt}.

\def\ZZ{C}
We only compute the retarded Green's function
$\GR^{tt}(r,r';p)$ for the operators
$j^t$ explicitly. The Green's function is equal to the retarded field
$j^t(r)$ induced by a small perturbation to the Hamiltonian
$\delta H=-j^t(r')e^{ip\cdot x}$.
That is to say $\GR^{tt}(r,r';p)$ solves
Maxwell's equation (with $A_r=0$ and $p_\mu=({-}\omega,0,0,k)$ for definiteness)
\begin{subequations}
\ba
\partial_r [\frac{\sqrt{{-}g} F^{tr}}{\g2}]
  + ik \frac{\sg F^{tz}}{\g2} &=& J^t, \label{maxwell2}
\\
  ik g^{zz} \partial_r A_z
    + i\omega |g^{tt}| \partial_r A_t
 &=& \frac{\g2}{\sg} J^r,
 \la{max1}
\ea
\end{subequations}
where the bulk current is
\label{src}
\ba
  J^t(r,x) &=&  -\frac{\sg |g^{tt}|}{\g2}(r') \,\partial_r\delta(r{-}r')
    e^{ip{\cdot}x}, \nl
  J^r(r,x) &=& -i\omega \frac{\sg |g^{tt}|}{\g2}(r') \,\delta(r{-}r')
    e^{ip{\cdot}x}, \nonumber
\ea
with $J^i=0$.
Note that the bulk current is conserved. By solving the second equation \nr{max1} for $\partial_r A_z$ and substituting
the result into a radial derivative of the first equation times $\ZZ$, we obtain a closed equation for $j^t(r;p)$ (and thus for $\GR^{tt}(r,r';p)$)
\be
\left[ \partial_r (\ZZ \partial_r)
  {+} (|g^{tt}|\omega^2{-}g^{zz}k^2)\ZZ\right]
 \GR^{tt}(r,r';p)
= -\partial_r \left(\ZZ \partial_r
 \frac{\delta(r{-}r')}{\ZZ g^{zz}}\right)
 - \omega^2\frac{|g^{tt}|}{g^{zz}}\delta(r{-}r').
\nonumber
\ee
where
\be
 \ZZ\equiv \frac{\g2}{\sg |g^{tt}| g^{zz}}. \nonumber
\ee

The second derivative of a delta function can be removed by adding
a contact term to the correlation function%
 \footnote{
  Such contact terms are generically present in correlation functions
  of $F^{\mu r}$.
  In flat space, for instance,
 $\GR^{tt}=
  {-}\frac{\partial_r^2+\omega^2}{\partial_r^2{+}\omega^2{-}k^2}
 \approx
 {-}\delta(r{-}r') {-}\frac{k^2}{\partial_r^2+\omega^2{-}k^2}$,
 which is structurally similar to \nr{eqlin}.
 }
\be
\left[ \partial_r (\ZZ \partial_r)
  {+} (|g^{tt}|\omega^2{-}g^{zz}k^2)\ZZ\right]
 \left[\GR^{tt}(r,r';p) + \frac{\delta(r{-}r')}{\ZZ g^{zz}} \right]
 = {-} k^2\delta(r{-}r').
 \la{eqlin}
\ee

The retarded function $\GR^{tt}$ obeys infalling boundary conditions
at the horizon, $\GR^{tt}\propto r^{-i\omega/2\pi T}$
and $\GR^{tt}\propto (r')^{-i\omega/2\pi T}$ \cite{sonstarinets}
and a Neumann condition $\partial_r \GR^{tt}=0$
at $r_c$. As seen from \nr{maxwell2},
the Neumann condition follows from the Dirichlet condition on $A_t$ and $A_z$.

The equation \nr{eqlin} is valid for any $\omega,k$.  We now solve it
in a hydrodynamical expansion.  We first solve the differential
equation \nr{eqlin} with vanishing right-hand side following closely
\cite{sonstarinets} and then patch the local solutions together to obtain
$\GR$. The near horizon behaviour can be factored out by writing
\be
 F(r)=e^{i\omega\int_r^{r_c} dr\sqrt{|g^{tt}|}} \,(1+\epsilon),
 \label{factorF} \nonumber
\ee
where $\epsilon$ goes to zero at the horizon. This is the only regular solution.
In the hydrodynamic regime, $\epsilon$ and its first derivative
are uniformly
small. The function $\epsilon$ solves the equation
\be
 \partial_r (\ZZ\partial_r\epsilon)
= i\omega \partial_r
\left(\ZZ\sqrt{|g^{tt}|}\right) + \ZZ g^{zz} k^2
+\OO(\omega\epsilon,k^2\epsilon).
 \label{eqe}
\ee
This can be integrated to give $\partial_r \epsilon$
\be
\partial_{r}\epsilon=i\omega(\sqrt{|g^{tt}|}-\frac{1}{\ZZ\sigma})
+\frac{1}{\ZZ}\int dr (\ZZ g^{zz}k^2)+\OO(\omega\epsilon,k^2\epsilon),
\ee
where $\sigma$ is an integration constant.
The regularity condition at the horizon implies  $\sigma = \lim_{r\to
  0}1/(\ZZ\sqrt{|g^{tt}|})$ .\footnote{This is the DC conductivity
of the plasma.}
One obtains
\be
 \partial_r F(r) = \frac{1}{\ZZ\sigma}\left[ -i\omega +
  \frac{\sigma k^2}{\Xi(r)}\right] +\OO(\omega^2\log\frac{1}{r},\omega k^2,k^4),
  \label{dummyparte}
\ee
where $\Xi^{-1}(r)=\int_0^r dr (Cg^{zz})$ has the interpretation
of a ``susceptibility'' at scale $r$ \cite{IqbalLiu}.
The $\omega\log \frac1{r}$ relative error signals the
breakdown of \nr{dummyparte} at exponentially small $r$
due to oscillations.
When $\omega\neq 0$, a second linearly independent solution is given
by $F^*(r)$. From \nr{dummyparte} the combination which obeys
Neumann boundary conditions at $r{=}r_c\to\infty$ is
\be
 G(r)=
  (1 - i\frac{\sigma k^2}{\omega \Xi}) F(r)
+ (1 + i\frac{\sigma k^2}{\omega \Xi}) F^*(r) \nonumber
\ee
where $\Xi=\lim_{r_c\to\infty} \Xi(r_c)$.
By gluing $G(r)$ at large $r$ to $F(r)$ at small $r$ and imposing
\nr{eqlin}, we finally obtain the bulk-to-bulk retarded function in the
hydrodynamic limit
\ba
 \GR^{tt}(r,r';p)+\frac{\delta(r{-}r')}{\ZZ(r) g^{zz}} &=& i\sigma k^2
\frac{ G(r)F(r')\theta(r{-}r') + F(r)G(r')\theta(r'{-}r) }
 {2(\omega+i\frac{\sigma}{\Xi}k^2)}
\nl
&=&
 \frac{i\sigma k^2}{\omega + i\frac{\sigma}{\Xi}k^2}
  +\OO(\omega\log\frac1{r},k^2).
 \label{GRtt}
\ea
Here the key observation is that in the hydrodynamic limit, the
bulk-to-bulk propagator $\GR^{tt}(r,r';p)$ loses its radial
dependence (after removal of a trivial $\delta$-function)
and becomes identical to the
boundary-to-boundary propagator. This can be traced back to
the $k^2$ factor on the right-hand side of \nr{eqlin} and smallness of the right-hand side in (\ref{radialflow}).

Equation \nr{GRtt} is in perfect agreement with the expectation
\nr{bulk2pt} with identifying
$D=\sigma/\Xi$ (known as Einstein relation).
We expect similar agreement for correlators of other fields.
Thus, as promised
at the beginning of this subsection, equations \nr{bulk2pt} hold
up to contact terms and higher order $\omega$ and $k$ corrections in the region $\mathcal{M}_{>}$, where the radii are
not exponentially small.
Bulk-bulk propagators other than the retarded one, follow from
the bulk KMS conditions \nr{KMSex}.
\subsection{Gauge-fixing, Faddeev-Popov ghosts and causality}\label{ghostsection}
At this point the reader might be puzzled that,
while we have emphasized on a causal formulation of the problem,
the gauge \nr{gauge1} we have chosen is \emph{acausal}. Causal gauges, such as light-cone, temporal or harmonic would have seemed more natural. On the other hand, the (spacelike) radial axial gauge is convenient so let us briefly pause to explain why this will not pose a real problem.

In order to gauge-fix, we introduce Faddeev-Popov ghosts.
Notice that the radial component
of the metric fluctuations, $h_{rr}$ and $h_{r\mu}$ are not fixed by
the Symanzik procedure; this only fixes the
induced metric on the boundary at $r=r_{c}$,
while one needs to impose (\ref{gauge1}) and (\ref{gauge2})
everywhere including in the bulk. The gauge fixing action we take is
\begin{equation}
 S_g= \int d^{d+1}x\sg \left(\lambda^{M}n^N h_{MN} + \lambda ~n{\cdot} A\right),
\end{equation}
where $n^{\mu}$ is the outward normal vector to the boundary. The fields $\lambda$ and $\lambda^{M}$ are Lagrange multipliers.
The ghost action associated to this gauge fixing is easily obtained
\begin{eqnarray}
 S_{FP}=\int d^{d+1}x \sg~\left[
 \tilde{c}^{M}n^N~\delta_\textrm{BRST} (g_{MN}) +
 \tilde{c}~\delta_\textrm{BRST} (n{\cdot}A) \right],
\end{eqnarray}
where
\ba
 \delta_\textrm{BRST} (g_{MN}) = \nabla_M c_N + \nabla_N c_M,\quad
 \delta_\textrm{BRST} (n{\cdot} A) = n{\cdot}\partial~c + A_M ~n{\cdot}\partial~c^M,
\ea
where $\nabla$ is the covariant derivative with respect to the full metric.
Note that the resulting ghost action is interacting and couples the gauge
and gravitational ghosts.

The ghost inverse propagator in the gravity sector is simply given by
\begin{equation}\label{ghostprop}
 i(G_M{}^N)^{-1}=\frac{\delta^{2}S_{FP}}{\delta \tilde{c}_{M}\delta
    c^{N}}= n\cdot\nabla^{(0)}\delta^{M}{}_{N}+n^M\nabla^{(0)}_{N}.
\end{equation}
Given that the orthonormal vector $n$ has only radial component,
it becomes clear that the ghost fields do not propagate in the field theory directions.
A similar behavior is found in the gauge sector. At vanishing chemical potential the gauge sector is decoupled.

A subtle issue here is that in order to
invert (\ref{ghostprop}), one needs to identify the boundary conditions
for the ghost fields.\footnote{
  Some of the boundary
  conditions can be deduced from the BRST invariance of the resulting
  gauge fixed action. For instance the BRST invariance implies that $c^{r}=0$
  on the boundary. This can be intuitively understood; the presence of the
  regularized boundary at some $r=r_c$ breaks gauge invariance in the
  radial direction; having a non-zero $c^{r}$ on the boundary will
  change the boundary so we must set it to zero.  In general,
  gauge-independence of physical observables should determine the
  boundary conditions.}


The unsettling issue seems to be that the ghosts itself can propagate on spacelike distances outside the causal diamond in Fig.~\ref{fig:kruskal}. This would seemingly place our claim that only quadrant $R$ is relevant to our formulation of the problem in danger.
It is a well-known fact that the spacelike axial gauge suffers from such problems \cite{Chan}. The gauge field propagators will also contain
``longitudinal parts'' that propagate along the radial direction and
similarly violate causality. From standard arguments, when computing an on-shell amplitude like what we
do here, the longitudinal modes will couple only among themselves (e.g., all
propagators in a loop will be simultaneously longitudinal). We expect that this will produce
a contribution of the same form which exactly cancels the acausal one from the ghosts in both
regions $\mathcal{M}_{<}$ and $\mathcal{M}_{>}$.
Let us recall that the amplitude we compute is gauge invariant.
One way to ``regularize'' the (spacelike) axial gauge (i.e., to make it causal) may
be to obtain it only as a limit of a causal gauge, similar to the proposal of \cite{Burnel}.
\subsection{Contribution from region $\mathcal{M}_<$}
\label{sec:proof}
Near the horizon the metric takes the Rindler form
\be
 ds^2= dr^2 - (2\pi r T)^2 dt^2 + \delta_{ij} dx^i dx^j.
\ee
The total volume of the half-almond (in the $r-t$ plane)
region $\mathcal{M_<}$ is of order $\epsilon^2$. With the choice \nr{epsilonchoice}, the volume is exponentially
small in $\tau$
\be
 V\propto \epsilon^2(\tau) \sim \frac{1}{T^2}e^{-\pi T\tau}.  \label{volumesuppression}
\ee
If the approximations which led to trivialized
the radial integrations in \nr{vertexg2} and \nr{vertexh2}
did hold in the small-$r$ region as well, computation
of the vertices would have been already completed. This is so because these radial integrations
would converge at their lower boundaries since the contribution from
$r< \epsilon$ scales like the volume \nr{volumesuppression}. However, we must be more cautious.
For instance, we have neglected $\partial_r A_i$ compared to $\partial_r
A_t$, yet we know from Gauss' law that
\be\label{GL}
 \partial_r A_z =  -\frac{|g^{tt}|}{g^{ii}} \frac{\omega}{k_z} \partial_r
 A_t \sim \frac{1}{r^2} \frac{\omega}{k_z} \partial_r A_t,
\ee
where $z$ is taken parallel to the loop momentum $k_z$
(which may, or may not, be perpendicular to the external indices $x$ and
$y$). As seen from (\ref{GL}), the factor $\omega/k_z$ gives a suppression away from the
horizon which justifies neglecting current fluctuations over charge fluctuation. But in the near horizon region the suppression does not exist anymore due to diverging metric factors. The second field in the loop, i.e., $g^{zz}\partial_r h^x{}_z$ in \nr{dummyggA}, satisfies an equation similar to (\ref{GL}). Using this, one can estimate the relative strength of a vertex involving current and stress in \nr{vertexg1} (a term which was neglected in $\mathcal{M}_{>}$) to that of involving charge and momentum fluctuations
\begin{eqnarray}
V_{A-A-g}&\supset&\int d^{d+1}x\sqrt{g} \delta{A_x}\frac{\delta F^{zr}}{g_{d+1}^2}\partial_{r}h^{x}_{~z}\sim (\frac{\omega}{k_z})^2\int d^{d+1}x\frac{1}{\sqrt{g}}\delta A_{x}j^{t}t^{t}_{~x}.
\end{eqnarray}
This is logarithmically divergent at small $r$,
since $j^t$ and $t^{t}_{~x}$ remain order 1 at small $r$.
The volume suppression (\ref{volumesuppression}) appears to be lost.
In the next subsection we argue that despite this fact, the contribution stemming from region $\mathcal{M}_{<}$ is vanishingly small.

%
\subsubsection{One vertex in $\mathcal{M}_<$ and one vertex in $\mathcal{M}_>$}\label{1-2}
In order to verify that the contribution from region $\mathcal{M}_{<}$ is small, we find it convenient to divide studying the radial integration for the one-loop amplitude, into two cases. First we concentrate on the case, where one of the vertices is located in $\mathcal{M}_{<}$ and the second resides in $\mathcal{M}_{>}$. Suppose that the lower vertex at $r_1$,
lies in $\mathcal{M}_<$, and the upper one at $r_2$ is located in $\mathcal{M}_>$.
In $\mathcal{M}_<$ all fields propagate along null geodesics. Furthermore \emph{both} fields in the loop must follow \emph{outgoing} null rays in order to connect with the exterior vertex at $r_2$.
Since both momenta are approximately null and collinear, one might expect that their Lorentz-invariant
dot product is much smaller than their individual components.
To be more precise, in the near horizon region using light-cone coordinates $x^\pm = t\pm r_{*}(r)$, one has
\be
  g^{MN}\partial_M \phi_1 \partial_N \phi_2
 \simeq g^{+-}\partial_+ \phi_1 \partial_- \phi_2 + (1\leftrightarrow 2)
 \ll g^{rr}\partial_r \phi_1 \partial_r \phi_2.
 \label{cancel}
\ee
Let us now check whether the vertices indeed take this form. In order to do so, we need some estimates.
In frequency space in the near-horizon limit, the possible
radial profiles for the various fields can
be readily found from the linearized equations of motion
(for the metric perturbations, see \cite{sonstarinetssound}). We find
\begin{align}
 h^{tt}&\sim r^{\pm i\omega}, r^{-3}, r^{-2}, &
 h^{tz}&\sim r^{\pm i\omega}, r^{-2}, 1, \nl
 h^{zz}&\sim r^{\pm i\omega}, r, r^2, &
 h^{ab}&\sim \delta{A}^x \sim r^{\pm i\omega}, \nl
 h^{at}&\sim \delta{A}^t \sim r^{\pm i\omega}, r^{-2}, &
 h^{az}&\sim \delta{A}^z \sim r^{\pm i\omega}, 1, \nonumber
\end{align}
Non-oscillatory modes in various channels
are pure gauge modes. These pure gauge modes are generated by the residual gauge symmetry after gauge-fixings (\ref{gauge1}) and (\ref{gauge2}) \cite{sonstarinetssound}. However, in section \nr{ghostsection} we have argued that the contribution from these pure gauge modes is canceled by ghosts loops, guaranteeing gauge invariance. Therefore, we should only worry about the ``physical''
part of the loop fluctuations
\be
 h^{\mu\nu} \sim r^{\pm i\omega},\quad \delta{A}^\mu \sim r^{\pm
   i\omega}
 \label{powercount}
\ee
where $\mu$ and $\nu$ are any space-time indices. We start from expression \nr{vertexg1} for the
gauge-gauge-graviton vertex
\be V_{A-A-g} = \int d^{d+1}x \,\frac{\delta A_x}{\g2}
 \partial_\alpha [ \sg(
    \delta F^{\mu\alpha}h^x_\mu
   +\delta F^{x\mu} h_\mu^\alpha
   -\frac12 \delta F^{x\alpha} h^\sigma_\sigma)].
\ee
By the estimates
\nr{powercount} the only dangerous terms near the horizon,
involve either two radial derivatives or two time derivatives
contracted together with an inverse metric $g^{tt}$.
We can neglect any explicit appearance of $g_{tt}$ so we drop any
field with a lower $t$ index. Such terms and terms
containing other type of derivatives are represented by $\OO(r^0)$  inside the bracket.
Using this it is easy to see that middle term in the bracket
is sub-leading. Going on-shell as before, in the near horizon limit, the other two
terms read
\be
 \label{cancel1}
 V_{A-A-g} \!\simeq\!\! \int \!\!d^{d+1}x \,\frac{\delta A_x}{\g2}\sg\!
\left[ -g^{MN} g^{ij}(\partial_M \delta A_i) (\partial_N h^x_j)
-\frac12 g^{MN} (\partial_M \delta A_x)(\partial_N h^i_i) +\OO(r^0)
\right]\!.
\ee
With the same power-counting we find that
the three-graviton vertex near the horizon is
\begin{eqnarray}
 V_{g-g-g}&=&{-}\frac{1}{8\pi G}\int d^{d+1}x \sg h^{xy} \delta^2 R_{xy}\\\nonumber
  &\simeq& \frac{-1}{8\pi G} \int d^{d+1}x \sg h^{xy}
 \left[ g^{MN}(\partial_M h^i{}_x) (\partial_N h_{yi})
              - g^{MN}\frac12 (\partial_M h_{xy})( \partial_N h^i{}_i)
+\OO(r^0)\right].
 \label{cancel2}
\end{eqnarray}
We see that both vertices \nr{cancel1} and \nr{cancel2} enjoy
the near-horizon cancelation \nr{cancel}; they vanish in the eikonal limit.
In order to estimate the sub-leading terms in the eikonal limit the following observation
is important. We can estimate the sub-leading corrections to the null-geodesic approximation of (\ref{powercount}). For a typical loop fluctuation
$\phi_i$ in the near horizon limit, we have $\phi_i\rightarrow r^{\pm
  i\omega}(1+\eta_i(r,\omega,k))$,
where $\eta\sim r^{\gamma_i}$ with $\Re \gamma_{i}>0$ in order to respect \nr{powercount}.
The correction $\eta$ is computable perturbatively.
If $\lambda=\min\{\Re \gamma_{i}\}$, from equations (\ref{cancel1})
and (\ref{cancel2}), one finds that the vertices are at least as small as
\begin{equation}
 V_{A-A-g}, V_{g-g-g}\sim [\epsilon(t)]^{\pm 2i \omega+\lambda}.
\end{equation}
The integration over them in $\mathcal{M}_<$ is thus
suppressed by $\epsilon^\lambda$ which is exponentially small in $\tau$.\footnote{Since
the integrand of a vertex is not gauge-invariant by itself,
for this argument to hold it is very important that the derivatives have
been organized not to act on the external $\delta{A}_x$ or $h^{y}_{~x}$ fields,
exactly as was mentioned in section (\ref{sec:vertex}).}
\subsubsection{Both vertices in $\mathcal{M}_<$}
There remains the case where both vertices reside in $\mathcal{M}_{<}$, where $r<\epsilon(\tau)$. The region $\mathcal{M}_{<}$ covers the ``tip neighborhood'' of the half-almond. We divide the discussion according to the spatial momentum circulating in the loop, small $k\lsim T$ or large $k\gg T$.

Consider a fluctuation with small spatial momentum $k$ running in the loop.
Since this region is flat, the relevant fields will follow either incoming or outgoing null geodesics.
Most importantly, there are no correlations between the two types. This is because correlations can only arise when the wave ``bounces''
against the geometry at large $r$ which takes a very long time (of order $\tau$).
Pick an incoming loop fluctuation. When it scatters against the incoming wave ($\delta{A}_x$ or $h_{xy}$), it will
remain purely incoming and un-deflected. Now when it is inserted into the second vertex, the cancelation
mechanism discussed previously will be effective and gives rise to the advertised suppression.
Recall that since both vertices are constrained to the region
$\mathcal{M}_<$, suppression of only one vertex is sufficient.
Diagrams with four-point vertices also exist in this region. They enjoy the same suppression.
This is because such a vertex will connect either to
three ingoing fields and one outgoing field or to three outgoing fields and one ingoing field. Since any contribution at large $r$
is exponentially small in $\tau$ (due to the external wavefunctions), one can freely employ integration-by-parts
to make all derivatives to act on the fields moving in the same direction.\footnote{
large derivatives of the metric will not be generated in this process, because the Rindler metric is flat.}

There remains only the large $k/T$ region. Manifestly this has little
to do with hydrodynamics. Since these modes decay very fast in time at large $r$, their contribution is
small asymptotically so we can freely perform integration-by-parts in them.
Since at large $k$ the exterior, curved region is separated by a large potential
barrier, we may think of this contribution as a gauge-invariant
one-loop scattering amplitude happening in just flat space.
This is an amplitude for a purely infalling perturbation
to \emph{reflect} back from the horizon into an outgoing fluctuation.
Note that, because the energies in the loop are exponentially large,
this may not necessarily be computable within
gravity; a stringy computation might be necessary. However, regardless of how it is computed, it appears
extremely unlikely that such an amplitude could be anything but
exponentially small (in $\tau$) -- a black hole event horizon is not supposed to
reflect perturbations even at quantum level.
A simple check of this claim is that it is true for any local
counterterm. In sum, although unlikely to be computable in pure gravity, it appears safe to ignore
the contribution from region $\mathcal{M}_{<}$.

This completes our argument that
with a natural choice for the action of the
derivatives in section \nr{sec:vertex}, only the region $\mathcal{M}_>$
is important.
\section{Assembling parts}\label{finalsection}

\subsection{Schwinger-Keldysh Basis}
\label{sec:pert}


It is most convenient to make a detour by first switching basis, within the two-dimensional vector space
in which the real-time fields $\phi_{1,2}$ ``live'', to the
so called Keldysh (or $(ra)$)\footnote{Here ``$r$'' meaning ``retarded''
 not to be confused with the radial coordinate. } basis.
The Keldysh basis fields consist of an averaged (or retarded) field
$\phi_r$ and a differenced (or advanced) field $\phi_a$
\be
 \phi_r=\frac{1}{2}(\phi_{1}+\phi_{2}),\quad \phi_{a}=\phi_{1}-\phi_{2}.
\ee
In this basis, the propagator is the $2\times 2$ matrix
\be
 G = \left( \begin{array}{ll}
      G_{rr} & G_{ra} \\ G_{ar} & G_{aa}
   \end{array} \right)
 =
 \left( \begin{array}{ll}
     {-}i(\GR{-}\GA) \left(\frac12{+}\nB(\omega)\right)
& {-}i\GR \\ {-}i\GA & 0
   \end{array} \right)
 \label{prop}
\ee
where we have used the equilibrium relations (\ref{KMS})
for the symmetric function
$G_{rr}=\frac12(G^>{+}G^<)$.

\begin{figure}[t]
\centerbox{0.50}{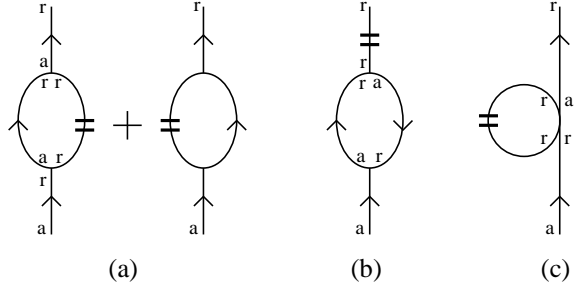}
\caption[Real-time Feynman diagrams]
{
 \label{fig:diags}
 (a) Keldysh index structure of the one-loop diagrams contributing to a retarded
     two-point function.
 (b) A diagram which vanishes due to a closed loop of retarded
     propagators.
 (c) A diagram with a four-point vertex (which is subleading in the
     long-time limit).
 Arrows indicate the flow of time along retarded propagators.
 The cut propagator is the $G_{rr}$ propagator (``symmetric propagator'').
}
\end{figure}

This basis is almost always the most efficient one to use.
Taking into account the fact that $G_{aa}=0$,
the only contributing configuration of the Keldysh indices for
a one-loop retarded amplitude is shown in Fig.~\ref{fig:diags}(a).
All interactions vertices carry an odd number of $\phi_a$
fields and we only need the vertices with exactly one $a$ index.

The physical meaning of this diagram should be quite transparent.
It amounts to expanding a retarded two-point function
to quadratic order in a background fluctuation, which is then averaged over an ensemble
with variance $G_{rr}$ (the doubly-dashed propagator in the figure).
This gives the full quantum-mechanical amplitude at one-loop.
The ``causality'' is manifest in from Fig.~\ref{fig:diags}(a); the products of retarded propagators
occur in such a way that both vertices must lie inside the causal diamond emanating from the
insertion points of the external operators. This is how the black hole interior region is manifestly avoided, as previously advertised.

\subsection{Long-time tails}

From the effective vertices \nr{vertexj} and \nr{vertext} and propagators
\nr{bulk2pt},
the agreement between the bulk computation and
the boundary hydrodynamics prediction is already
obvious, since the radial coordinate has now disappeared from the problem.
The remaining steps of the computation are nonetheless tedious.
We combine \nr{vertexj}, \nr{bulk2pt} and the
diagrammatic results of the preceding subsection to obtain the
one-loop retarded correlator for $j^x$.
At this stage we can safely go back to frequency
space, since the radial integrations have been taken care of,
alleviating the need of manifest space-time causality:
\ba
 {}^{(1)}\GR^{xx}(\omega,k{=}0) &=&
  \frac{2T}{w^2}\int_k 
  \int \frac{d\omega'}{2\pi}  \nl
&&
 \hspace{-1.5cm} \left[
  \frac{\Im \GR^{tt}(\omega',k)}{\omega'} \GR^{tx,tx}(\omega{-}\omega',k)
   + \GR^{tt}(\omega',k) \frac{\Im \GR^{tx,tx}(\omega{-}\omega',k)}{\omega{-}\omega'}
 \right],
\ea
where $w=\epsilon+p$. We have used $G_{rr}\simeq (2T/\omega) \Im \GR$
in the loop, which is valid at small frequencies $\omega,\omega' \ll T$.
The integrations are straightforward but slightly tedious
\ba {}^{(1)}\GR^{xx}(\omega,k{=}0) &\simeq&
  i\frac{2T \Xi D \gamma_\eta}{w}\int_{k} k^2(k_y^2{+}k_z^2)
  \int \frac{d\omega'}{2\pi} \nl
&& \hspace{-2cm}
\left[
 \frac{1}{(\omega'{+}iDk^2)[(\omega{-}\omega')^2+(\gamma_\eta k^2)^2]}
 +\frac{1}{(\omega'{}^2 +(Dk^2)^2)(\omega-\omega'+i\gamma_\eta k^2)}\right]
 \nl
&=&
  \frac{T \Xi D \gamma_\eta}{w}\int_{k}
 \frac{D{+}\gamma_\eta}{D\gamma_\eta}
\frac{k_y^2{+}k_z^2}{\omega+i(D{+}\gamma_\eta)k^2}
\nl
&\simeq& \frac{T\Xi}{2w} (d{-}2)\Gamma(\frac{1-d}{2})
\left(\frac{-i\omega}{4\pi(D{+}\gamma_\eta)} \right)^{\frac{d-1}{2}},\quad \omega>0.
 \label{finalJ}
\ea
When $\omega<0$ we take the complex conjugate.
Note on the second line we have dropped the sound
contribution from $\GR^{tx,tx}$, which does
not contribute at large times since its pole does not pinch with the
diffusion one in $\GR^{tt}$ \cite{KovtunYaffe}.
Similarly, on the last line we have kept only the term which is non-analytic at $\omega=0$;
this originates from small $k$ and is not affected by the fact that the $k$ integration is ultraviolet-divergent.

The approximations made in this paper make it impossible to
discuss the low-frequency
regime of $G$, only the long-time regime, e.g. the non-analytic behavior
at small $\omega$.

Taking the imaginary part of \nr{finalJ} and multiplying it by $2T/\omega$ gives the
fluctuation function. We write it in time domain
\begin{eqnarray}
{}^{(1)}G_{rr}^{xx}(t) &=&
 \frac{T^2\Xi}{w} \frac{d{-}2}{(4\pi(D{+}\gamma_\eta))^{\frac{d-1}{2}}}
  \Gamma(\frac{1-d}{2})\sin(\frac{\pi}{2}\frac{1-d}{2})
 \int_{-\infty}^\infty \frac{d\omega}{2\pi} e^{-i\omega t} |\omega|^{\frac{d-3}{2}}
\nl
&=&
 \frac{T^2\Xi}{w} \frac{d-2}{d-1} \frac{1}{[4\pi
     (D{+}\gamma_\eta)|t|]^{\frac{d-1}{2}}}
 +\OO(\frac{1}{t^{\frac{d-1}{2} +1}}),
\end{eqnarray}
which exactly reproduces the hydrodynamic prediction \nr{hydro1}.
Note that this is the same as $\int d^{d-1}x ~{}^{(1)}\!\langle \{ j^x(t,x),j^x(0)\} \rangle$.

For $t^{xy}$ the calculation is similar but it uses the effective
vertex \nr{vertext}, $t^{xy}\simeq t^{xt}t^{yt}/(\epsilon{+}p)$.
By now it is easy to commute the use of fluctuation-dissipation relations with the Fourier
transforms\footnote{
 Equation \nr{resT} is only exact in the soft approximation
 $\nB\approx T/\omega$. In general, with obvious notation,
 the correct formula for a $(rr)$ self-energy is $\Pi_{rr}=\int \frac12 (G^>G^> + G^<G^<)$,
 where $G^{>,<}$ are Wightman functions.
},
making the equivalence with the hydrodynamics computation in section
\nr{hydro}, even more clear
\ba
{}^{(1)}G_{rr}^{xy,xy}(t,k{=}0) &=&
 T^2 \int_k \left[G_{rr}^{tx,tx}(t,k) G_{rr}^{ty,ty}(t,k) +
	          G_{rr}^{tx,ty}(t,k) G_{rr}^{ty,tx}(t,k)\right]
 \label{resT}
\nl
 = \textrm{Eq.}~(\ref{hydro2}).
\ea
The \emph{bulk} symmetric functions $G_{rr}$ in \nr{bulk2pt}
appear here. They are precisely equal to the boundary ones \nr{2ptfunc}.
\section{Concluding remarks} \label{sec:conc}
So far, in the context of AdS/CFT, the \emph{phenomena} associated to the
non-linear hydrodynamics, had mostly remained unexplored. In this presentation,
we have shown how to reproduce hydrodynamic
long-time tails generated by non-linearities
from a one-loop AdS gravity computation. Universality of the hydrodynamic description was clearly observed within gravity computation. Only very basic entries of the AdS/CFT dictionary were used.

Based on scale/radius relation in AdS/CFT, one might have expected
that the long-time tails originate from the region
in the immediate vicinity of the black hole event horizon,
where the gravitational red-shift factors are large.
In our setup, it turns out that this expectation is not entirely true. It is true that the very existence of the long-lived diffusive fluctuations is intimately tied to the near horizon region, however we find that the scale at which these fluctuations interact, is
set by another scale. Indeed, the interaction vertices, from a bulk point of view, localize at the same
radius at which susceptibilities receive most of their contributions
(see the integral \nr{vertexg2}), i.e., the scale $T$.

\begin{figure}[t]
\centerbox{0.750}{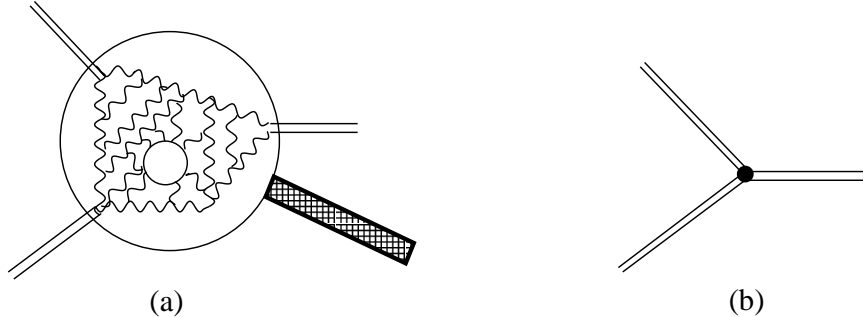}
\caption[N=4 SYM processes contributing to a hydrodynamic vertex]
{
\label{fig:magnifying}
(a) N=4 SYM microscopic processes contributing to (b) a
 hydrodynamic vertex.  In the regime considered in this paper, (b) is
also equivalent to a gravity vertex.}
\end{figure}

The way to reconcile this finding with the scale/radius correspondence is to
observe that the interactions are associated with the microscopic
scale $T$ and not the long-time scale $\tau$. In fact, in hydrodynamics,
the interactions are exactly instantaneous (see Fig.~\ref{fig:magnifying}).
If the interactions had been localized in the near horizon region, they
would have corresponded to nonlocal (in time) interactions from the
boundary theory viewpoint. This could not have been reconciled
with the hydrodynamics predictions.

From a subset of our computation, contained in subsection \ref{sec:vertex},
one can deduce the interaction between three boundary hydrodynamic modes.
The interactions found from the bulk gravity are in agreement with
the hydrodynamic prediction. Related computations have appeared before, to all orders in nonlinearity in
hydrodynamics \cite{Shiraz}.  Our method is complimentary.
We have shown that the leading bulk gravity
Feynman diagrams with three external legs reproduce the
leading non-linear interactions in hydrodynamics.

As expected, the ``correct'' degrees of freedom to use in the bulk are the
(radial) momenta conjugate to the metric and gauge field. This is nice
because the boundary limit of the (radial) conjugate momenta are
exactly the boundary charges and
currents which are ``the variables'' in hydrodynamics.
These variables describe fluctuations of the horizon.
In the hydrodynamic regime, they obey simple radial flow equations \nr{radialflow}.

The one-loop computation appearing in this paper is technically novel, given its dynamical and real-time nature, although a different one-loop AdS computation,
in the context of AdS/condensed matter physics,
has appeared prior to this work in \cite{HD1,HD2}.

It is not evident from our computation whether the asymptotically AdS region is necessary for the long-time tail phenomenon to occur.
It would be interesting to look for similar phenomena
close to the horizon of black holes in asymptotically flat space,
in suitable correlators and on suitable time scales.

In the presence of continuous phase transitions,
hydrodynamic description must include the order parameter itself as
a degree of freedom surviving the long-wavelength limit.
Including its effect might provide an interesting
extension of this work.

\section*{Acknowledgments}

We are grateful to Laurence G. Yaffe, Sean Hartnoll and Pavel Kovtun for reading the manuscript. We would like to thank Alex Maloney, Guy Moore and Simon Ross for discussions.
Both authors are supported by the NSERC of Canada.


\end{document}